\documentclass[prd,aps,preprint,amsmath,nofootinbib,amssymb,eqsecnum,showkeys]{revtex4-1}
\pdfoutput=1
\usepackage{verbatim,graphics,graphicx,color,slashed,textcomp}
\usepackage{ulem} 
\usepackage[colorlinks=true,
            linkcolor=red,
            urlcolor=blue,
            citecolor=blue]{hyperref}

\graphicspath{{figures/}}

\newcommand{\GeV}{\mathrm{GeV}}

\begin{document}

\title{Interacting Scalar Radiation and Dark Matter in Cosmology}
\author{Yong Tang}
\email{ytang@kias.re.kr}
\affiliation{Korea Institute for Advanced Study, \\
	85 Hoegiro, Dongdaemun-gu, Seoul 02455, South Korea}
\date{\today}

\begin{abstract}
We investigate possible cosmological effects of interacting scalar radiation and dark matter. After its decoupling, scalar radiation can stream freely as neutrinos or self-interact strongly as perfect fluid, highly depending on the magnitude of its self-couplings. We obtain the general and novel structure for self-scattering rate and compare it with the expansion rate of our Universe. If its trilinear/cubic coupling is non-zero, scalar radiation can be eventually treated as perfect fluid. Possible effects on CMB are also discussed. When this scalar also mediates interaction among dark matter particles, the linear matter power spectrum for large scale structure can be modified differently from other models. We propose to use Debye shielding to avoid the singularity appearing in the scattering between scalar radiation and dark matter. 
\end{abstract}
\maketitle

\section{Introduction}\label{sec:intro}
According to our current understanding, nearly $95\%$ of energy density in our universe consists of dark components, namely dark energy and dark matter. The standard cosmological model, a cosmological constant with cold dark matter, called $\Lambda$CDM, is very successful at large scales~\cite{Planck:2015xua}. At small scales, there are controversies that allow scenarios beyond collisionless CDM, see Ref.~\cite{Weinberg:2013aya} for a recent review.

Although not all of these dark components are necessarily connected, it should not be very surprising that some could have new interactions. If dark matter has significant interactions beyond gravitation, there could be dramatically different predictions that can be tested by observations. For instance, when a light particle mediates the interaction between dark matter, we can get enhanced annihilation cross section, a possible scenario for positron fraction excess in cosmic ray data~\cite{,ArkaniHamed:2008qn}\footnote{The excess can also be explained by models in which DM in scenario with non-standard cosmology and interactions can have enhanced perturbation at small scales~\cite{Choi:2015yma}. More substructures or subhalos could arise and give a large boost factor. }. If dark matter has large self-interaction, its density distribution around galactic center tends to have a flat profile~\cite{Spergel:1999mh}. If dark matter interacts with some relativistic particle in cosmic background, matter power spectrum could get suppressed~\cite{Boehm:2000gq, Boehm:2004th, Green:2005fa, Loeb:2005pm, Bertschinger:2006nq, Bringmann:2006mu}, relaxing the ``missing satellite'' problem~\cite{Moore:1999nt, Kravtsov:2009gi}. 

There are various models in particle physics that can provide the above mentioned interaction. DM with gauge or global symmetries is extensively discussed in~\cite{Aarssen:2012fx, Bringmann:2013vra, Chu:2014lja, Dasgupta:2013zpn, Ko:2014bka, Ko:2014nha, Nelson:2014mva, Chu:2015ipa, Arhrib:2015dez, Kang:2015aqa, Kainulainen:2015sva, Bernal:2015bla, Choi:2015bya, Ma:2015roa, Chacko:2015noa, Baek:2013dwa, Buen-Abad:2015ova, Heikinheimo:2015kra}. Atomic and mirror DM can also have similar phenomenology~\cite{Cline:2013zca, Cline:2013pca, Foot:2004wz, CyrRacine:2012fz}. Different DM models within supersymmetric framework are explored in~\cite{ Boddy:2014yra, Kang:2016xrm}. Closely related model-independent analyses about effects on astrophysics are conducted in~\cite{Feng:2009hw, Buckley:2009in, Loeb:2010gj, Tulin:2013teo, Petraki:2014uza, Buckley:2014hja, DelNobile:2015uua, ethos, Bernal:2015ova, Binder:2016pnr}. 

In this paper, we investigate a new, illustrating model with interacting scalar radiation and dark matter, and discuss the possible cosmological effects on cosmic microwave background (CMB) and large scale structure (LSS). Scalars can have cubic and quartic self-interactions, which can affect their cosmological evolution. If these interactions are small enough, scalar radiation is streaming freely after decoupling and behaves just as neutrinos. If these interactions are not negligible, scalar may be treated as perfect fluid and affects CMB differently. The interaction between dark matter and scalar radiation also induces novel temperature dependences in scattering cross section, which are crucial in cosmological context and lead to imprints on linear power spectrum.

This paper is organized as follows. In Sec.~\ref{sec:model} we set the theoretical framework by introducing the explicit model. Then in Sec.~\ref{sec:cmb} we investigate how scalar contributes as radiation by changing the effective number of neutrinos, whether it streams freely or behaves as perfect fluid, and what the possible effects on CMB. Next in Sec.~\ref{sec:lss}, we consider the cosmological effects of scattering between DM and scalar radiation. We propose to use Debye shielding to avoid the singularity appearing in the scattering process. Finally, we give our conclusion.

\section{Interacting Scalar Radiation and Dark Matter}\label{sec:model}
We start with the very simple but general Lagrangian density with (pseudo-)scalars $\phi_i$ and fermionic dark matter $\psi$,
\begin{equation}\label{eq:lag}
\mathcal{L}=\mathcal{L}_{\textrm{SM}}+ \bar{\psi}(i\slashed{\partial}-m_\psi)\psi -\bar{\psi}(g^s_i+ig^p_i \gamma_5)\psi \phi_i+ \frac{1}{2}\partial_\mu \phi_i \partial^{\mu} \phi_i -\mathcal{V}\left(\phi_i, H\right),
\end{equation}
where $m_\psi$ is the mass of $\psi$, $g^s_i$ and $g^p_i$ are respectively the scalar and pseudo-scalar type coupling constants, and $H$ is the standard model Higgs doublet. Repeated index is summed. We have introduced a set of scalars $\phi_i$ for reasons which we shall discuss shortly. It is surprising that the above Lagrangian has not been discussed in the cosmological context. As we shall show in this paper, such a simple model has some novel features and interesting implications. If dark matter is a scalar field $X$, we can study the phenomenology of $X$ and $\phi_i$ similarly by introducing interaction terms like $
X^\dagger X \left( \mu_{i}\phi_i + g_{ij}\phi_i\phi_j\right).$

Scalars $\phi_i$ can be massive or massless, and the resulting cosmology could be quite different. Throughout our discussion, we shall not specify the fundamental origins of these scalars. In particle physics, scalars are ubiquous, such as Higgs field, axion, inflation field, bound states, dark energy, and so on. Without loss of generality, we may first discuss the $\phi$ part in the potential $\mathcal{V}$,
\begin{equation}\label{eq:potential}
	\mathcal{V}\left(\phi_i, H\right)\supset \frac{1}{2}m^2_i \phi_i^2 + \frac{\mu_{ijk}}{3!}\phi_i \phi_j \phi_k + \frac{\lambda_{ijkl}}{4!}\phi_i \phi_j \phi_k \phi_l, 
\end{equation}
where the individual $m_i,\mu_{ijk}$ and $\lambda_{ijkl}$ can be zero or non-zero. Introduction of interaction terms with $H^\dagger H$ has at least one immediate effects that $\psi$ and $\phi_i$ can be thermalized in the early Universe. To have the correct electroweak vacuum, the potential need satisfy some conditions, see Appendix for detailed discussion. It is also possible to introduce other non-renormalizable terms like $\frac{1}{\Lambda}\bar{\psi}\psi H^\dagger H$ to do the thermalization.

The relic density of DM $\psi$ is basically determined by the couplings to $\phi_i$. $\psi$ can be produced either through usual thermal freeze-out or freeze-in process, it can also produced by heavy $\phi$'s decay. If $\psi$ and all $\phi_i$ are heavy, say heavier than $\GeV$, phenomenologies in these aspects are the same as traditional cold dark matter and it makes no difference in our model, Eq.~\ref{eq:lag}.

However, if there is a light state in $\phi_i$, although the relic density calculation is probably only modified by including Sommerfeld effects~\cite{Hisano:2003ec, Hisano:2004ds}, there are other very important consequences on cosmological observables, such as CMB and large scale structure (LSS), which are the main topics in this paper. As we shall show that the details not only depend on the interaction between DM $\psi$ and $\phi_i$, but also on self-interaction terms $\mu_{ijk}$ and $\lambda_{ijkl}$. 

\begin{figure}[t]
	\includegraphics[width=0.8\textwidth,keepaspectratio]{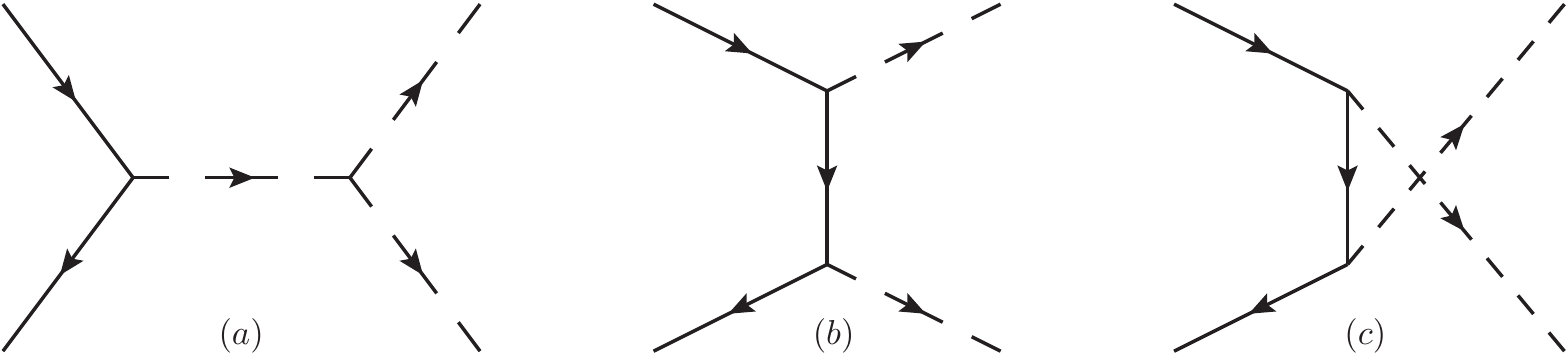}
	\caption{\label{fig:thermal} Thermal processes for $\bar{\psi}+\psi \leftrightarrow \phi_i + \phi_j$. Here and after, solid and dashed lines represent fermion $\psi$ and scalar $\phi_i$, respectively.}
\end{figure}

\section{Scalar Radiation and CMB}\label{sec:cmb}
Assume $\phi_1$ is the light state, massless or having a very tiny mass compared with its temperature $\left(m_{\phi_1}\ll T_{\phi_1}\right)$, one immediate effect is that $\phi_1$ will contribute as radiation in cosmic background. The convenient quantity to account for this contribution is the effective number of neutrino species, $N_{\textrm{eff}}$, which describes how much relativistic species are present in our Universe. $\phi_1$ would increase $N_{\textrm{eff}}$ by
\begin{align}\label{eq:geff}
\delta  N^{\phi_1}_{\textrm{eff}}&\equiv \frac{\rho_{\phi_1}}{\rho_{\nu}}=\frac{4}{7}\frac{T^4_{\phi_1}}{T^4_{\nu}}
=\frac{4}{7}\left[\frac{g_{\ast s}\left(T_\nu\right)}{g_{\ast s}^{\phi}\left(T_{\phi_1}\right)} \times 
\frac{g_{\ast s}^{\phi}\left(T_{\phi_1}\right)T_{\phi_1}^{3}}{g_{\ast s}\left(T_\nu\right)T_{\nu}^{3}}\right]^{\frac{4}{3}}\nonumber\\
&= \frac{4}{7}\left[ \frac{g_{\ast s}\left(T_\nu\right)}{g_{\ast s}^{\phi}\left(T_{\phi_1}\right)}
\frac{g_{\ast s}^{\phi} \left(T^{\textrm{dec}}\right)(T^{\textrm{dec}})^3}{g_{\ast s}\left(T^{\textrm{dec}}\right)(T^{\textrm{dec}})^3}
\right]^{\frac{4}{3}}
= \frac{4}{7}\left[ \frac{g_{\ast s}\left(T_\nu\right)}{g_{\ast s}^{\phi}\left(T_{\phi_1}\right)}
\frac{g_{\ast s}^{\phi} \left(T^{\textrm{dec}}\right)}{g_{\ast s}\left(T^{\textrm{dec}}\right)}
\right]^{\frac{4}{3}},
\end{align}
where $T$ is the temperature, $g_{\ast s}$ counts the effective degrees of freedom for entropy density in standard model sector, or particles that are in kinetic equilibrium with neutrinos, while $g_{\ast s}^{\phi}$ denotes the effective degrees of freedom that are in kinetic equilibrium with $\phi_1$. And we have used entropy conservation in the last equality. Although the exact value depends on the kinetic decoupling temperature $T^{\textrm{dec}}$ and ratios of degrees of freedom before and after decoupling, the typical value for $\delta  N_{\textrm{eff}}$ would be around $\mathcal{O}\left(0.1\right)$ which is definitely allowed by present data~\cite{Planck:2015xua, Zhang:2014ifa, Huang:2015wrx}. For instance, if $T^{\textrm{dec}}\sim 1\GeV$, we have $\delta  N_{\textrm{eff}}\simeq 0.045$. If more than one scalar contributes as radiation, we should rescale $\delta  N_{\textrm{eff}}$ correspondingly. The exact value of $T^{\textrm{dec}}$ is determined by the interaction with standard model particle. Simple calculation shows that an interaction term $\lambda_{\phi H}\phi_1^2 H^\dagger H$ with $\lambda_{\phi H}\sim 10^{-3}$ would give $T^{\textrm{dec}}\sim 1\GeV$. 

We shall note there could be other relativistic particles that are in kinetic equilibrium with $\phi_1$, which would also contribute to extra $\delta N_{\textrm{eff}}$, and $\delta  N_{\textrm{eff}}$ is then changing with time. For example, in the paper we considered $T^{\textrm{dec}}\sim 1$GeV. If the dark matter $\psi$ is lighter than $1$GeV, $\psi$ could be still in kinetic equilibrium with $\phi_1$ and would contribute to $\delta N_{eff}$. After its decoupling from $\phi_1$ at $m_{\psi}/25$, $\psi$ would transfer its entropy to $\phi_1$ and $\phi_1$'s temperature is effectively increased. This is all encoded in counting $g_{\ast s}^{\phi}\left(T_{\phi_1}\right)$ in the above formula, Eq.~\ref{eq:geff}.

We know in standard model neutrinos are decoupled after BBN time and then start {\it free-streaming}, which means the interactions of neutrinos can be neglected so that perturbations in its anisotropic stress and high multipole can develop. However, in our model $\phi_1$ is not necessarily free-streaming after its kinetic decoupling from standard model thermal bath and it may self-scatter a lot and acts like a perfect fluid that has no anisotropy and high multipole. Whether and when $\phi_1$ is streaming freely  depends crucially on its self-couplings or interaction with other relativistic particles. 

\begin{figure}[t]
	\includegraphics[width=0.95\textwidth,keepaspectratio]{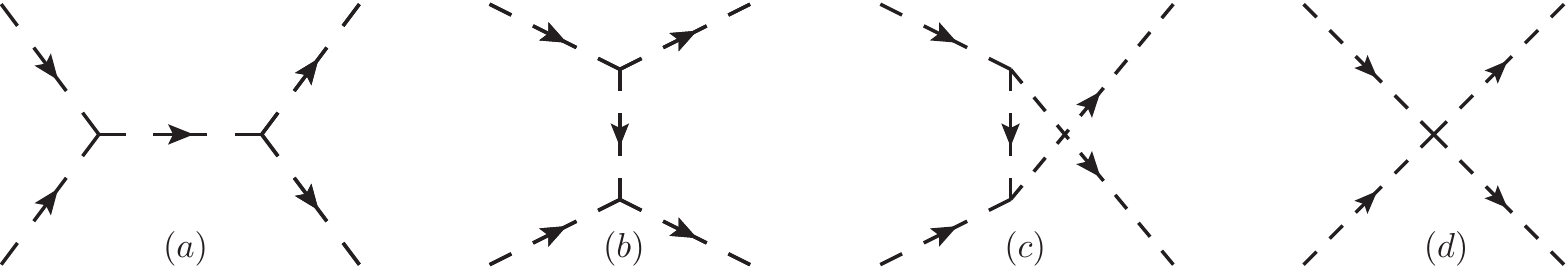}
	\caption{\label{fig:self} Feynman diagrams for self-scattering of $\phi_1 + \phi_1 \rightarrow \phi_1 + \phi_1$. Contributions from the first three diagrams are proportional to $\mu_1^4$ and from the last one are proportional to $\lambda_1^2$.}
\end{figure}

The self-scattering rate of $\phi_1$ is dominantly determined by $\phi_1 + \phi_1 \rightarrow \phi_1 + \phi_1$ through the Feynman diagrams shown in Fig.~\ref{fig:self}. Define $\mu_{1}\equiv \mu_{111}$ and $\lambda_{1} \equiv \lambda_{1111}$, we can estimate the scattering rate as
\begin{equation}\label{eq:gself}
	\Gamma_{\phi_1}= n_{\phi_1}\times\langle \sigma v \rangle \sim  T^3_{\phi_1}\times\left[\frac{3\mu^4_1}{T_{\phi_1}^6}+\frac{\lambda^2_1}{T_{\phi_1}^2}\right]
	=\frac{3\mu^4_1}{T_{\phi_1}^3}+\lambda^2_1 T_{\phi_1},
\end{equation} 
where we have neglected some numeric factors $\sim\mathcal{O}(0.1-10)$ which are not essential for illustrating the main physical effects. Contributions from the first three diagrams in Fig.~\ref{fig:self} are proportional to $\mu_1^4$ and from the last one are proportional to $\lambda_1^2$.

The most important features of Eq.~\ref{eq:gself} are the temperature dependences in comparison with the evolution of Universe. $\lambda$-term in Eq.~\ref{eq:gself} with linear temperature dependence can also be obtained from other interactions, for instance, fermionic radiation with gauge interactions. While the $\mu$-term with inverse cubic power law, as far as we know, is not presented elsewhere. 

\begin{figure}[t]
	\includegraphics[width=0.5\textwidth,keepaspectratio]{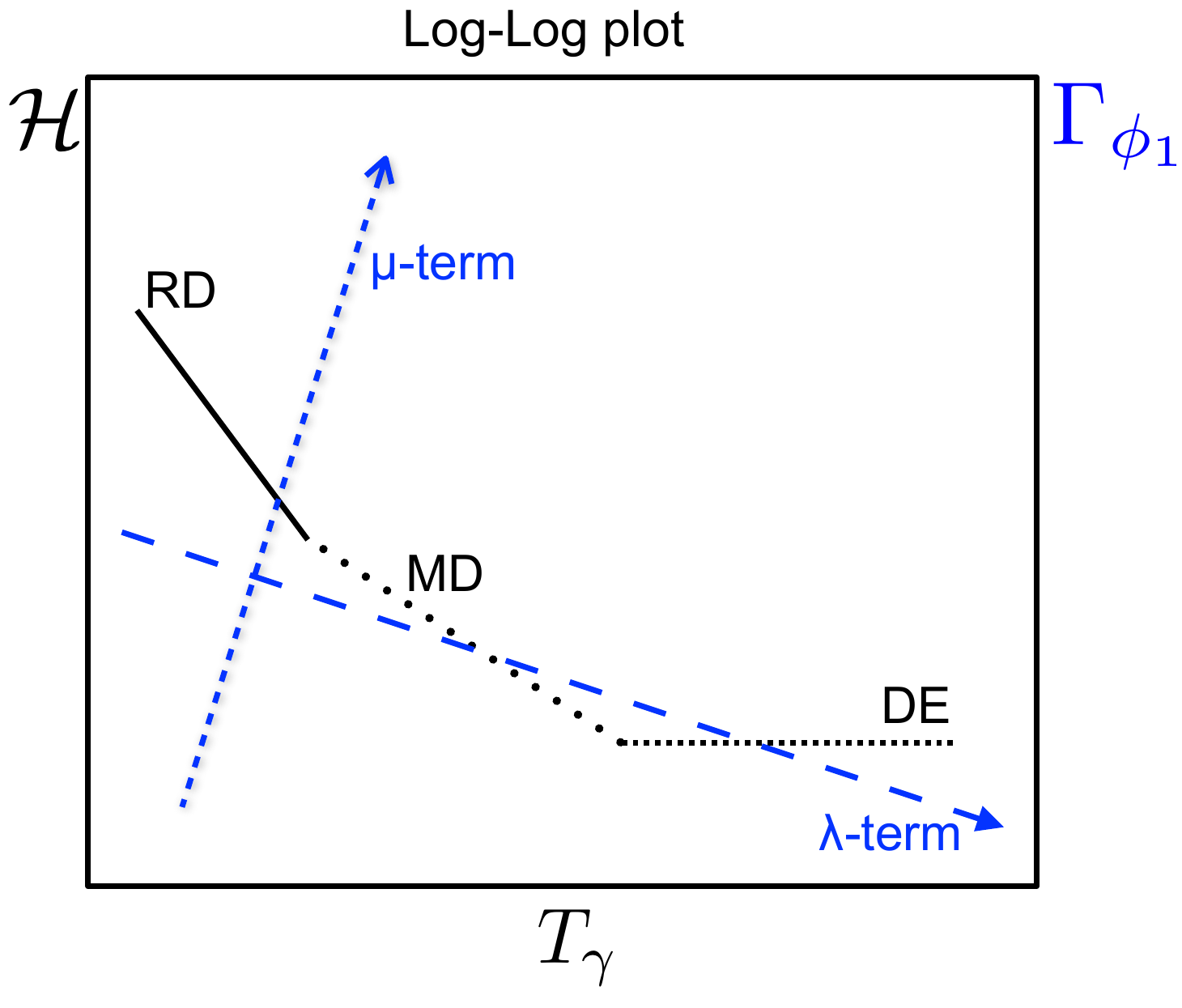}
	\caption{\label{fig:Hubble} Schematic plot for $\mathcal{H}$ and $\Gamma_{\phi}$ as photon temperature $T_\gamma$ decreases. Black lines, solid, dotted, and dashed ones respectively show $\mathcal{H}$ in radiation, matter and dark energy dominant times. Evolutions of $\mu$-term and $\lambda$-term in $\Gamma_{\phi_1}$, Eq.~\ref{eq:gself}, are shown in blue median-dashed line and long-dashed line, respectively. Increasing or decreasing $\mu_1$ and $\lambda_1$ will shift upwards or downwards globally.}
\end{figure}

Recall that the expansion rate or typical time scale in cosmic evolution with flat spatial curvature is set by Hubble parameter, $\mathcal{H}$,
\begin{equation}
	\mathcal{H}^2\equiv \left(\frac{\dot{a}}{a}\right)^2 = \frac{8\pi G}{3}\sum_i \rho _i
	\Rightarrow  \mathcal{H}=\sqrt{\frac{8\pi G}{3}}\left[ \rho_{r0}\left(\frac{T_{\gamma}}{T_{\gamma 0}}\right)^4+\rho_{m0}\left(\frac{T_{\gamma}}{T_{\gamma 0}}\right)^3+\rho_{de}\right]^{1/2},
\end{equation}
where we have the energy density for radiation $\rho_r$, matter $\rho_m$ and dark energy or cosmological constant $\rho_{de}$,  $a$ is the scale factor ($a_0=1$ for present value), $G$ is Newton's constant and quantities with subscript `0' stand for the present values.

Note that $\mathcal{H}$ has a different temperature dependence from $\Gamma_{\phi_1}$, therefore $\mathcal{H}/\Gamma_{\phi_{1}}$ is changing as temperature goes down or as Universe expends. Since for $\delta  N_{\textrm{eff}}\sim\mathcal{O}\left(0.1\right)$ we have $T_{\phi_1}\sim 0.5 T_{\gamma}$, the energy density of $\phi_1$ is just about one order-of-magnitude smaller than photon and we would expect several interesting cases could happen, depending on how large $\mu_1$ and $\lambda_1$ are. These cases are
\begin{enumerate}
\item { $\mu_1 \neq 0$}: There must be a time at which $\Gamma_{\phi_1}\gtrsim \mathcal{H}$. Once crossing this time point, $\phi_1$ can be treated approximately as a perfect relativistic fluid.

\item {$\mu_1=0$ but $\lambda_1\neq 0$}: Whether $\phi_1$'s self-scattering is important or not crucially depends on the size of $\lambda_1$. But if $\Gamma_{\phi_1}\gtrsim \mathcal{H}$ happens, it can only be reached  first at radiation or matter dominate era. In dark energy dominant epoch $\Gamma_{\phi_1}$ will be eventually smaller than $\mathcal{H}$ since $\Gamma_{\phi_1}$ is decreasing but $\mathcal{H}$ is nearly constant. 

\item {$\mu_1=0$ and $\lambda_1 = 0$}: $\phi_1$ is streaming freely after its kinetic decoupling just like neutrinos in standard cosmology.

\end{enumerate}

The above discussion can be best illustrated with a schematic plot in Fig.~\ref{fig:Hubble}, where $\mathcal{H}$ and $\Gamma_{\phi_1}$ are shown as functions of photon temperature $T_\gamma$ in log-scale. As $T_\gamma$ decreases towards to the right-hand side, $\mathcal{H}$ experiences first radiation dominant (RD) era as black solid line, then through the matter dominant (MD) epoch shown in dotted line, and finally dark energy (DE) dominant time with dashed line.  Evolutions of $\mu$-term and $\lambda$-term in $\Gamma_{\phi_1}$, Eq.~\ref{eq:gself}, are shown in blue median-dashed line and  long-dashed line with arrow, respectively. Increasing or decreasing $\mu_1$ and $\lambda_1$ will shift the corresponding arrowed line upwards or downwards globally. All the above mentioned cases can be understood by shifting the arrowed lines. 

\begin{figure}[t]
	\includegraphics[width=0.7\textwidth,keepaspectratio]{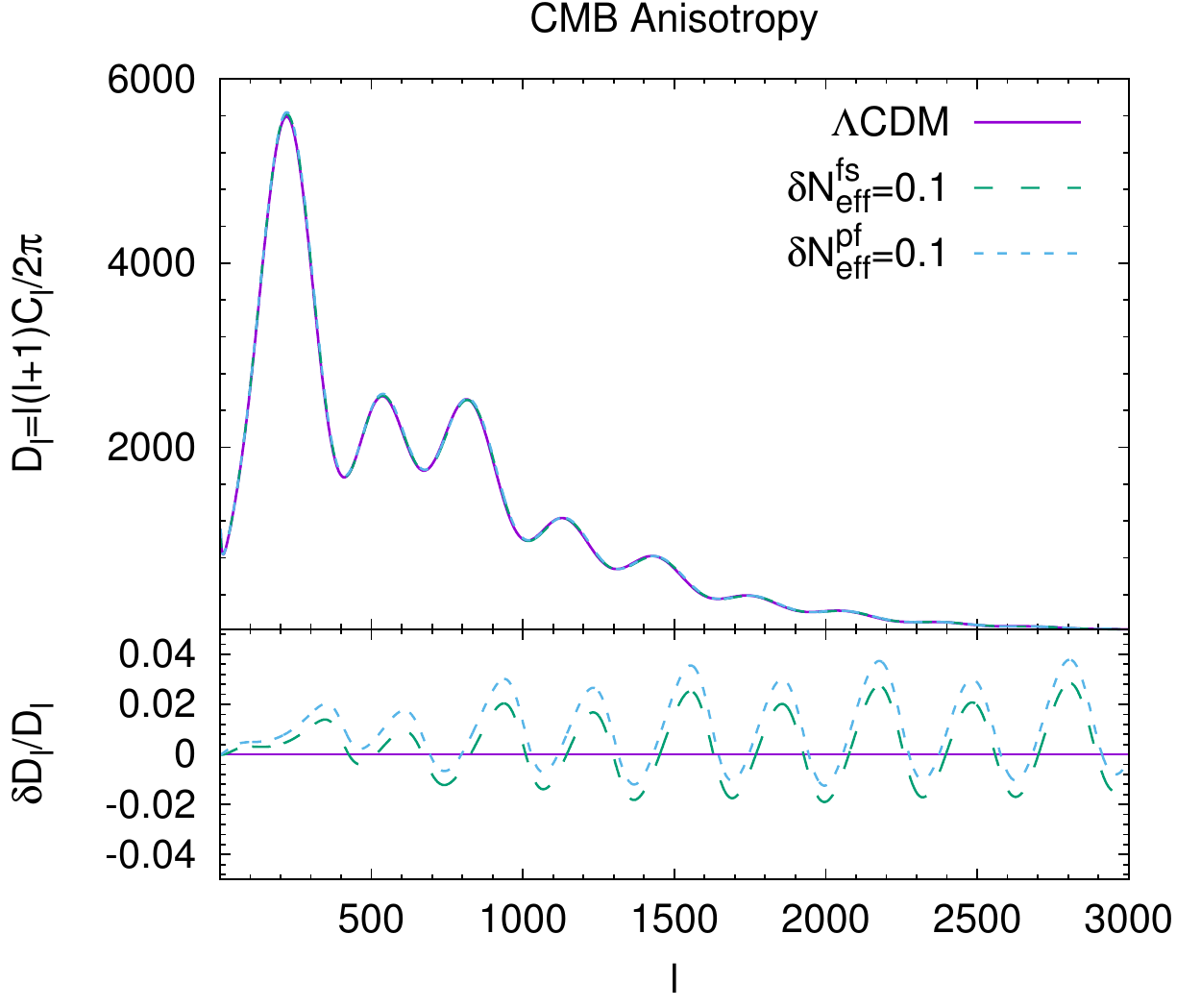}
	\caption{\label{fig:cl} Effects of $\delta N_{\textrm{eff}}=0.1$ on CMB temperature anisotropy. Dashed (Long-dashed) line corresponds to the case with perfect fluid (free-streaming) radiation, as shown in the upper plot for the overall effect. In the lower plot, we show the relative difference from the standard $\Lambda$CDM, at order of $\mathcal{O}(1\%)$. See text for details.}
\end{figure}

In Fig.~\ref{fig:cl}, we show the effects $\delta N_{\textrm{eff}}=0.1$ on CMB temperature anisotropy. Dashed (Long-dashed) line corresponds to the case with perfect fluid (free-streaming) radiation. When calculating the power spectrum, we have modified the public Boltzmann code {\tt{CLASS-2.4.3}}~\cite{class}. For the perfect fluid case, the maximal effect arise when radiation starts to behave as perfect fluid in radiation dominate era. In other instances, it will lie in the middle of perfect fluid and free-streaming. In the upper plot, we see that it is almost indistinguishable from the standard $\Lambda$CDM due to relative small difference at order of $\mathcal{O}(1\%)$, shown in the lower plot. It is expected that in future CMB experiment high precision measurement would be able to resolve the difference. 

\section{Interacting Dark Matter and LSS}\label{sec:lss}

\begin{figure}[t]
	\includegraphics[width=0.8\textwidth,keepaspectratio]{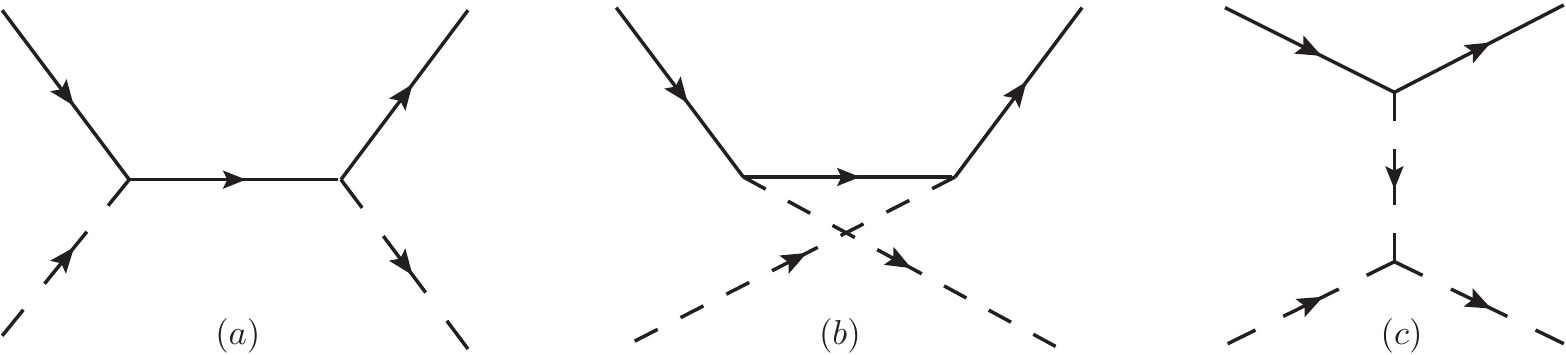}
	\caption{\label{fig:scatter} Elastic scattering for $\psi + \phi_1 \rightarrow \psi + \phi_1$, $\psi$ and $\phi_1$ are displayed as solid and dashed lines, respectively. The first two diagrams, (a) and (b), mimic the Compton scattering, $e^-+\gamma \rightarrow e^- +\gamma$, which the last one (c) gives some unique features and has no equivalent in other models, see text for details. }
\end{figure}

In this section, we shall investigate cosmological effects of the interaction between DM $\psi$ and scalar radiation $\phi_i$. The overall relic density of DM $\psi$ has already been discussed in Sec.~\ref{sec:model}. Here we shall only focus on the effects from elastic scattering between DM and $\phi_1$.

Before $\psi$'s kinetic decoupling, elastic scattering, $\psi + \phi_1 \rightarrow \psi + \phi_1$, keeps $\psi$ in equilibrium with $\phi_1$. The scattering rate is given by the calculation of Feynman diagrams in Fig.~\ref{fig:scatter}. The first two diagrams, (a) and (b), mimic the Compton scattering, $e^-+\gamma \rightarrow e^- +\gamma$, while the last one (c) gives some unique features and has no equivalent in other models. 

The scattering rate of contributions from (a) and (b) in Fig.~\ref{fig:scatter} have the following behavior,
\begin{equation}\label{eq:kineticab}
	\Gamma^{a+b}_{\psi\phi_1}=n_{\phi_1}\langle \sigma v \rangle_{a+b}\sim T^3_{\phi_1}\times\left[\frac{\left(g_1^{s}\right)^4}{m^2_\psi}
	+\frac{\left(g_1^{p}\right)^4}{m^4_\psi} T^2_{\phi_1}\right],
\end{equation}
where the first term in bracket is constant, similar to Thomson scattering limit, and the second term depends on temperature quadratically. Such temperature dependences are typically studied~\cite{Boehm:2014ksa} for DM models with vector or scalar mediator. In our considered model, Eq.~\ref{eq:lag}, we have an additional diagram,  Fig.~\ref{fig:scatter}(c), which as we shall show below, has totally different behavior.

The contribution from the last one, (c), is given by
\begin{equation}\label{eq:kineticc}
\Gamma^c_{\psi\phi_1}=n_{\phi_1}\langle \sigma v \rangle_{c}\sim T^3_{\phi_1}\times\left[\frac{\left(g_1^{s}\right)^2 \mu^2_1}{T^4_{\phi_1}} + \frac{\left(g_1^{p}\right)^2\mu_1^2}{m^2_\psi} \frac{1}{T^2_{\phi_1}}\right],
\end{equation}
where we see again that inverse power-law arises, due to exchanging of $\phi_1$, similar to $\mu$-term in Eq.~\ref{eq:gself}. When obtaining the above equations, we have neglected some numeric factors which again do not affect qualitatively the physical effects. We also notice that scalar and pseudoscalar interactions have different temperature dependence in cosmological evolution. For scalar dark matter, we do not have the $g_1^{p}$-term in Eq.~\ref{eq:kineticab} and Eq.~\ref{eq:kineticc}. 

\begin{figure}[t]
	\includegraphics[width=0.7\textwidth,keepaspectratio]{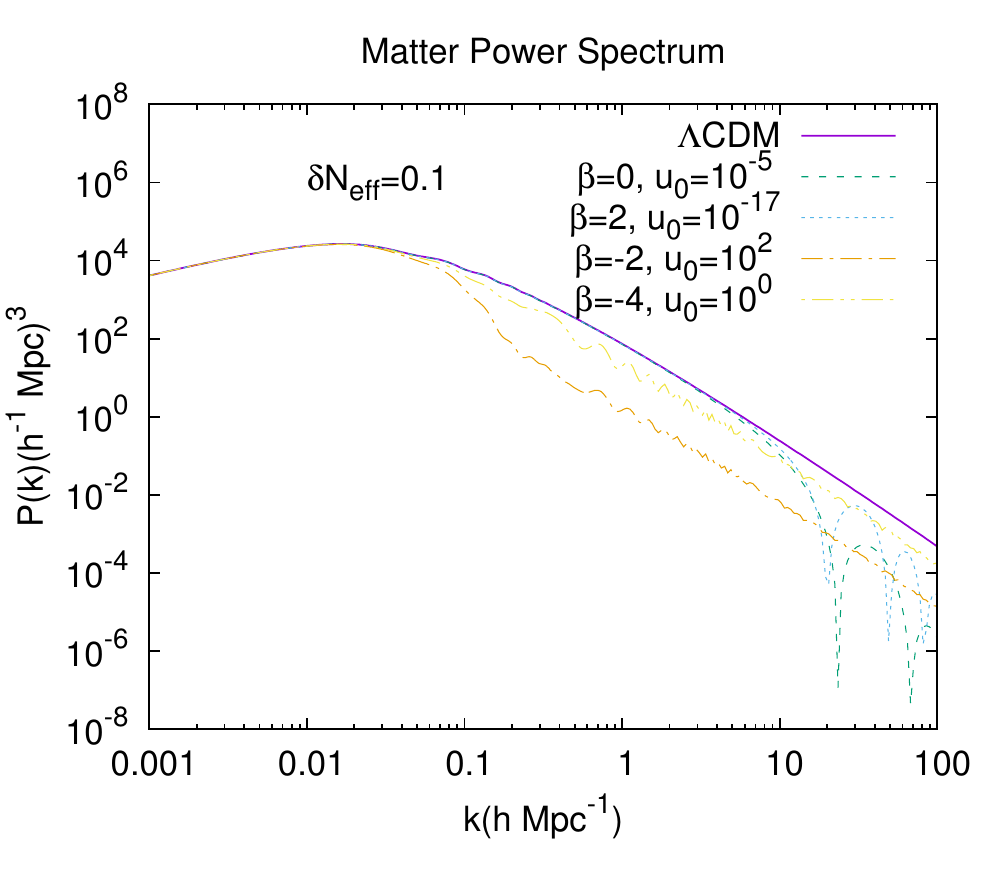}
	\caption{\label{fig:matterpower}Illustration of matter power spectra in cases with different temperature dependences. Parameters are defined in Eq.~\ref{eq:parametrize}. See text for details. }
\end{figure}

The cosmological effects from the elastic scattering is that collisional damping will be induced on matter power spectrum. Typically, oscillation behavior should arise, similar to baryonic acoustic oscillation.  The momentum relaxation rate~\cite{Green:2005fa, Loeb:2005pm, Bertschinger:2006nq, Bringmann:2006mu} determining the kinetic decoupling of $\psi$-$\phi_1$ is given by
\[
\gamma\left(T_{\psi}\right)\equiv\frac{T_{\psi}}{m_{\psi}}\times\Gamma_{\mathrm{\psi \phi_1}},
\]
and should be compared with Hubble parameter $\mathcal{H}$. If $\gamma\left(T_{\psi}\right)$ decreases more quickly than $\mathcal{H}$ does as shown in Eq.~\ref{eq:kineticab}, then at lower temperature kinetic decoupling happens when $\gamma\left(T_{\psi}\right) \lesssim \mathcal{H}\left(T_{\mathrm{kd}}\right)$. The corresponding collisional damping scale manifests itself as a characteristic scale in the matter power spectrum with 
\begin{equation}
M_{\textrm{c}}=\frac{4\pi}{3}\rho_{\textrm{M}}\left(\frac{1}{\mathcal{H}\left(T_{\mathrm{kd}}\right)}\right)^{3}\sim 2\times10^{8}\left(\frac{T_{\textrm{kd}}}{\textrm{keV}}\right)^{-3}M_{\odot},
\end{equation}
where $\rho_{\textrm{M}}$ is the sum of matter densities, $\rho_{\textrm{CDM}}+\rho_{\textrm{baryon}}$. Below this scale, the power spectrum is suppressed, see Fig.~\ref{fig:matterpower} for a quick glimpse. Interestingly, if $M_{\textrm{c}}\sim\mathcal{O}(10^{9})M_{\odot}$, it might be able to resolve the ``missing satellites" problem.

However, if $\Gamma^c_{\psi\phi_1}$ or $\mu_1$ in Eq.~\ref{eq:kineticc} is non-zero, $\gamma/\mathcal{H}$ can actually increasing as our universe cools down due to the inverse power-law temperature dependence. In such a case, matter power spectrum at very large scale could also be affected. To show quantitative results, we need solve the cosmological perturbation evolutions for $\psi$ and $\phi_1$. Explicitly, similar to photon-baryon system~\cite{Ma:1995ey}, the Euler equation for DM and $\phi_1$ would be modified to 
\begin{align}
	\dot{\theta}_{\phi_1} &=k^2 \Psi +k^2\left(\frac{1}{4}\delta_{\phi_1} -\sigma_{\phi_1}\right)-\Gamma_{\psi}\left(\theta_{\phi_1}-\theta_{\psi}\right),\\
	\dot{\theta}_\psi &= k^2\Psi -H \theta_\psi + R^{-1}\Gamma_{\psi}\left(\theta_{\phi_1}-\theta_{\psi}\right),
\end{align}
where dot means derivative over conformal time $\tau$ defined by $\textrm{d}t=a\textrm{d}\tau$,  $\theta_{\phi_1}$ and $\theta_\psi$ are scalar $\phi_1$ and DM $\psi$'s velocity divergences, $k$ is the comoving wavenumber, $\Psi$ is the gravitational potential, $\delta_{\phi_1}$ and $\sigma_{\phi_1}$ are the density perturbation and anisotropic stress potential of $\phi_1$, and $H$ is the conformal Hubble parameter, $\dot{a}/a$, the interaction rate $\Gamma_{\psi}=a n_{\psi}\sigma_{\psi \phi_1}c $ and the density ratio $R=\frac{3}{4}\rho_\psi/\rho_{\phi_1}$. We implement the above equations into the public Boltzmann code {\tt CLASS-2.4.3}~\cite{class} and approximately treat $\phi_1$ as perfect fluid with $\sigma_{\phi_1}\simeq 0$.

Now we parametrize the cross section ratio at the current temperature
\begin{equation}\label{eq:parametrize}
u_0 \equiv \left[\frac{\sigma_{\psi \phi_1}}{\sigma_{\textrm{Th}}}\right]\left[\frac{100\GeV}{m_\psi}\right], u_\beta(T)=u_0 \left(\frac{T}{T_0}\right)^\beta,
\end{equation}
where $\sigma_{\textrm{Th}}$ is the Thomson cross section, $0.67\times 10^{-24}$cm$^{-2}$. $\beta = -4,-2,0, 2$ correspond to individual terms in Eqs.~\ref{eq:kineticab} and \ref{eq:kineticc}. 

With $\delta N_{\textrm{eff}}=0.1$ we illustrate different cases in Fig.~\ref{fig:matterpower}. The solid line corresponds the $\Lambda$CDM cosmology which other lines are labeled with different $\beta$ and $u_0$ defined in Eq.~\ref{eq:parametrize}. With positive $\beta$, the matter power spectra are affected more in the large $k$ or small scales, while negative $\beta$ can modify also very large scales or small $k$ region. All these behaviors are expected as we explain above under Eqs.~\ref{eq:kineticab} and \ref{eq:kineticc}. To make quantitative constraints on the cross section, one need to conduct Markov chain Monte Carlo analysis, which is beyond our scope here. 

There is one more important feature we would like to take a close look. When calculating the differential scattering cross section in Fig.~\ref{fig:scatter}(c) for $\psi + \phi_1 \rightarrow \psi + \phi_1$, we have
\begin{equation}
\frac{d\sigma}{d\Omega}=\frac{1}{64\pi^{2}s}\left|\mathcal{M}\right|^{2},
\end{equation}
with $s\simeq m_\psi^2$, $d\Omega=2\pi \textrm{d}\cos{\theta}$, $\theta$ is the scattering angle, $\left|\mathcal{M}\right|^{2}$ is the matrix element,
\begin{equation}
\left|\mathcal{M}\right|^{2} \simeq  \mu^2_1\left[\frac{\left(g_1^{s}\right)^2}{p^4_{\phi_1}\left(1-\cos{\theta}\right)^2} + \frac{\left(g_1^{p}\right)^2}{m^2_\psi p^2_{\phi_1}\left(1-\cos{\theta}\right)}\right].
\end{equation}
To obtain Eq.~\ref{eq:kineticc}, we have taken the thermal replacement approximation, $p^2_{\phi_1}\rightarrow T^2_{\phi_1}$, and $\left(1-\cos{\theta}\right)\sim 1$. However, there is actually singularity near $\cos{\theta}=1$, which is the infrared divergence, common in quantum field theories with massless particles. $\theta=0$ means zero-momentum transfer or infinite long range interaction but without scattering. So, this singular region have no effect on $\psi- \phi_1$ scattering, similar to Rutherford scattering in quantum electrodynamics. For our purpose here, one straightforward way to get rid of the singularity is just to introduce a small mass for $\phi_1$ so that there is a finite length beyond which the interaction is effectively vanishing. Other way is to regularize the integration through
\begin{equation}
\int d\Omega \left[\left(1-\cos{\theta}\right)^2\frac{d\sigma}{d\Omega}\right],
\end{equation}
which is finite now. 

Here we propose another way to circumvent the singular problem, motivated by plasma physics. We introduce one more scalar $\phi_2$ which has couplings to $\psi$ with a relative sign difference from $\phi_1$'s, namely
\begin{equation}\label{eq:sign}
g^s_2/ g^s_1 < 0,g^p_2/g^p_1 < 0.
\end{equation}  
When interaction between $\phi_1$ and $\phi_2$ is attractive, $\phi_1$ is then surrounded with $\phi_2$s which effectively shield the interaction between $\phi_1$ and $\psi$, similar to a phenomena called Debye shielding in electromagnetic plasma. The corresponding Debye length in our model is estimated as
\begin{equation}
\lambda_D\sim \left(\frac{T^3_{\phi_1}}{n_{\phi_1}\left(g_1^{s}\right)^2\mu_{221}^2}\right)^{1/2},
\end{equation}
where $\mu_{221}$ is the coupling for vertex $\phi_2\phi_2\phi_1$ in scalar potential Eq.~\ref{eq:potential}. This length corresponds to a minimal momentum transfer $\delta p^2_{\textrm{min}}=p^2_{\phi_1}(1-\cos{\theta_{\textrm{min}}})$, equivalently a minimal scattering angle $\theta_{\textrm{min}}$. Therefore, singularity can also be avoided.

\section{Conclusion}
In this paper, we have investigate some plausible cosmological effects from interacting scalar radiation and dark matter (DM). After its kinetic decoupling, scalar radiation can be streaming freely as standard neutrino or interacting strongly as perfect fluid, which leads to distinguishable effects on cosmic microwave background. When scalar radiation can be treated as perfect fluid depends on the self-interaction strength. If its trilinear or cubic term is non-vanishing, massless scalar eventually behaves as perfect fluid. 

The frequent scattering between DM and scalar radiation leaves imprint on the matter power spectrum which is an important probe in large scale structure. This scattering can decay the kinetic decoupling of DM and give rise to collisional damping or oscillation in DM density perturbation, similar to baryonic acoustic oscillation. The suppression in matter power spectrum might be test with future experiment. We also identify a novel structure of temperature dependence in the scattering amplitude where singularity appears. We propose use Deybe shielding to avoid this singular problem.

\begin{acknowledgments}
The author would like to thank Celine Boehm, Qing-Guo Huang and Xin Zhang for enlightening discussions, and Ryan Wilkinson for helps with Boltzmann code on related project. This work is partly supported by National Research Foundation of Korea Research Grant NRF-2015R1A2A1A05001869.
\end{acknowledgments}

\section*{Appendix}
Here we discuss the general conditions for the potential $\mathcal{V}$ to give the correct electroweak vacuum. Let us take the potential for the radiation field $\phi$ and standard model doublet Higgs field $H$ as
\begin{eqnarray*}
	\mathcal{V}\left(\phi,H\right) &\supset  & V\left(\phi\right) + \left( \mu_\phi \phi + \lambda_{\phi H}\phi^2\right)  H^\dagger H -\mu^2_{H} H^\dagger H + \lambda_H \left(H^\dagger H\right)^2,\nonumber \\
	V\left(\phi\right) &=&\lambda_1 \phi + \lambda_2 \phi^2 + \lambda_3 \phi^3 + \lambda_4 \phi^4, 
\end{eqnarray*}
To have the correct electroweak vacuum, $\langle H\rangle=v_H/\sqrt{2},v_H\simeq 246$GeV, we can impose the following minimum conditions, positivity and stability at infinity, 
\begin{align*}
\left. \frac{\partial \mathcal{V}}{\partial \phi}\right|_{\phi=0}  = 0 & \Rightarrow \lambda_1 = -\mu_\phi v_H^2/2, \; \\
\left. \frac{\partial^2 \mathcal{V}}{\partial^2 \phi}\right|_{\phi=0}  >  0 & \Rightarrow \lambda_2 > -\lambda_{\phi H} v_H^2/2,\; \\
\left. \mathcal{V} \right|_{\phi,H\rightarrow \infty}  > 0 & \Rightarrow \lambda_4 > 0,\lambda_H>0 \textrm{ and } \lambda_{\phi H}> -2\sqrt{\lambda_4\lambda_H} ,\\
\mathrm{Det} \mathcal{M}^2 &=
\mathrm{Det} \left(
\begin{array}{cc}\label{eq:mass}
2\lambda_2+\lambda_{\phi H}v^2_H & \mu_\phi v_H/2 \\
\mu_\phi v_H/2 & 2\lambda_H v^2_H
\end{array}
\right) >0,
\end{align*}
and the condition that $\phi=0$ is the global minimum leads to 
\begin{equation*}
\lambda_2+\lambda_{\phi H} v^2_H/2+ \lambda_3 \phi + \lambda_4 \phi^2 >0\Rightarrow
\lambda^2_3-4\lambda_4\left(\lambda_2+\lambda_{\phi H} v^2_H/2\right)<0. 
\end{equation*}
These constraints would maintain the correct electroweak vacuum. As shown in above formulas, the masses of $\phi$ and Higgs are the eigenvalues of the mass matrix. Even if we take the mixing parameter $\mu_\phi=0$, the mass of $\phi$ is $2\lambda_2+\lambda_{\phi H}v^2_H$. So to have very light $\phi$, we would need the $\lambda_2$ to cancel out the large mass contribution from higgs field. This indeed could lead theoretical issues, like fine tunning.

\providecommand{\href}[2]{#2}\begingroup\raggedright\endgroup

\end{document}